\documentclass[10pt]{amsart}
\usepackage[]{graphicx}
  \graphicspath{{figures/}}
  \usepackage{color}

\usepackage[]{amsmath}

\begin{document}
%
\title{Superposition Principle in Linear Networks with Controlled Sources}
\author{Ciro~Visone}
\thanks{C. Visone is with the Department
of Engineering, Universit\`a degli Studi del Sannio, Benevento Italy, e-mail: (visone@unisannio.it)}

\begin{abstract}
The manuscript discusses a well-known issue that, despite its fundamental role in basic electric circuit theory, seems to be tackled without the needful attention. The question if the Principle of Superposition (POS) can be applied to linear networks containing linear dependent sources still appears as an addressed point unworthy to be further discussed. Conversely, the analysis of this point has been recently re-proposed \cite{M-L-94, Damper_11} and an alternative conclusion has been drawn. From this result, the manuscript provides an alternative approach to such issue from a more general point of view. It is oriented to clarify the issue from the didactic viewpoint, rather than provide a more efficient general technique for circuit analysis.
By starting from a linear system of equations, representing a general linear circuit containing controlled sources, the correct interpretation of ``turning off'' the controlled elements in terms of \emph{circuit equivalent} is provided, so allowing a statement of the POS for linear circuits in a wider context. Further, this approach is sufficiently intuitive and straightforward to fit the needs of a Basic Electric Circuit Theory class.
\end{abstract}

\maketitle

\textbf{keywords} - Circuit theory; Linear circuits; Superposition; Linear controlled sources (LCS);

\section{Introduction}
Everyone, during his studies learned to exploit the Superposition Principle as an effective tool in a large number of subjects, all sharing the linearity assumption. In Circuit Theory, superposition takes a fundamental role both as operational and `conceptual' simplification. Each of us has been taught that in the circuit we are allowed to turn off \emph{only} the independent sources and all classical handbooks, such as \cite{Chua}-\cite{Nilsson} seem to assume such a viewpoint.\\
However, some different (and sound) opinion have arisen in the past years. In \cite{M-L-94} W.M. Leach described his experience on the subject and discussed the possibility to apply the Superposition Principle also to linear circuits containing linear controlled sources. Such a paper is well-written and very instructive and addressed a very basic issue perhaps tackled without sufficient attention. R. I. Damper, in \cite{Damper_11} renewed the subject and strengthen Leach conclusion. A further and interesting contribution to this field was provided by T. S. Rathore, et. al in \cite{Rathore_12}, who discussed the issues proposed in Leach and Damper contributions, evidencing a link between the Surposition Principle approach and the Miller equivalents, \cite{Rathore_2010_1}, and evidencing an increase of the computation efficiency when the so-called Matrix Approach is exploited, \cite{Rathore_2010_2}.

In the following, taking all the above contributions in mind, a different approach is proposed. It starts from the analysis of the algebraic system describing the circuit model, underlining that moving a term with an unknown variable to the right hand side of the equation has an interesting circuit interpretation that can be fruitfully exploited by students and practicing electrical and electronic engineers. So the manuscript, rather than propose more efficient techniques, is rather aimed to show the possibility to extend the Superposition Principle also to linear circuits containing linear \emph{controlled sources}, by evidencing that all the tools available for the analysis of linear circuit by the aid of the Superposition Principle still hold and can be fruitfully applied.

\section{The general model of a resistive linear circuit}

\begin{figure}[hb]
\centering
\includegraphics[width=6cm]{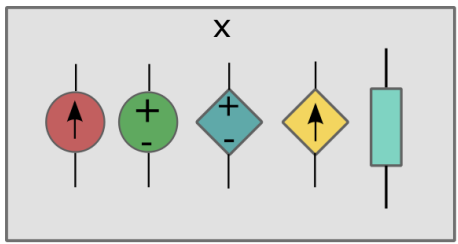}
\caption{Linear resistive circuit with linear resistors, independent and controlled voltage and current sources. The $x$ variable represents the vector of current and voltage circuit unknowns.}
\label{fig:general_circuit}
\end{figure}

The motivation that the application of the superposition tool is forbidden when dependent sources are concerned lies in the misleading interpretation that a controlled source cannot be really turned off, from the circuit side. This in turn derives from the intuitive persuasion that a dependent source cannot provide any voltage or current when independent sources are \emph{switched off}. However, even if this is correct from the physical viewpoint, it seems to forget that the analysis of physical systems is almost ordinarily carried out by \emph{formal} tools without a specific `counterpart' in the \emph{physical world}, \cite{Damper_11}.\\
To this aim, let us consider a linear circuit with $b$ branches and $n$ nodes, containing linear resistors, independent and dependent sources, as sketched in Fig. \ref{fig:general_circuit}.
%
Applying the Kirchhoff's Laws and the constitutive relationships for the elements, the general form of the algebraic system of equations modeling such a circuit is as follows:
\begin{eqnarray}
\mathrm{A}\mathbf{i}  = 0; \label{Tableau_1} \\
\mathrm{B}\mathbf{v} = 0; \label{Tableau_2} \\
\mathrm{M}\mathbf{v} + \mathrm{N}\mathbf{i} = 0, \label{Tableau_3}
\end{eqnarray}
being $\mathrm{A}$ the $(n-1)\times b$ \emph{incidence} matrix, $\mathrm{B}$ the $[b-(n-1)]\times b$ \emph{fundamental loop} matrix, while $\mathrm{M}$ and $\mathrm{N}$ are $b\times b$ \emph{coefficient} matrices specifying the constitutive relationship of all network elements. Further, $\mathbf{i}$ and $\mathbf{v}$ are $b\times 1$ vectors of currents and voltages, respectively. 

The $(2b\times 2b)$ linear algebraic system, specified by eqns. \eqref{Tableau_1}-\eqref{Tableau_3} can be written as 
\begin{equation}\label{eq:algebra_1}
\mathcal{L} x = U,
\end{equation}
where, \emph{for the sake of example}, could assume the form:\\
\begin{equation}\label{eq:algebra_2}
\mathcal{L} = \begin{bmatrix}
 & \textbf{\large A} & & &  & \textbf{\large 0} &  \\
 & & & & & \\
 & \textbf{\large 0} & & & & \textbf{\large B} & \\
\hdotsfor{7} \\
 0 & \dots  0& -R \hdots & & 0 & \hdots & 1 \hdots 0 \\
0 & 0 &\hdots & & 0 & \dots 1 & 0 \hdots\\
\hdotsfor{7} \\
0 & \hdots  & 1  \hdots & & -y_{2b-1} & 0 \hdots & 0 \hdots \\
0 & -y_{2b} & \dots 0 & & 0 & 0 \hdots  & 1 \hdots \\
\end{bmatrix}.
\end{equation}
\\
This is a \emph{nonsingular} $(2b\times 2b)$ matrix and $U$ is a $(2b\times 1)$ vector with nonzero elements in the same rows where the independent sources of the $\mathcal{L}$ matrix are located. Of course, since $\mathcal{L}$ is nonsingular and equations are independent by hypothesis, the system has only one solution. \\
%
\begin{figure}[hb]
\centering
\includegraphics[width=7cm]{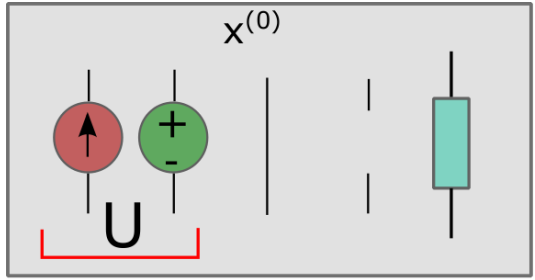}
\caption{Sub-circuit where only the independent sources are turned on.}
\label{fig:general_circuit_2}
\end{figure}
%
\begin{figure}[hb]
\centering
\includegraphics[width=7cm]{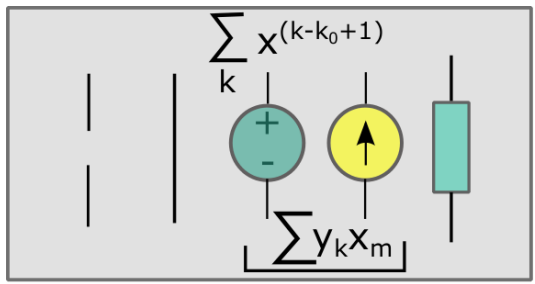}
\caption{Sub-circuit where the dependent sources play the role as if they were independent sources. The real independent sources are turned on.}
\label{fig:general_circuit_3}
\end{figure}
\\
Let us further assume that each row from $b+1$ to $k_0-1$ ($k_0>b+1$) contains either only a $+1$ element (this corresponds to the constitutive characteristic of an independent voltage or current source), or at least two nonzero elements $L_{i j}$ and $L_{i m}$, $i = b+1, ...,k_0-1$ such that their distance is $|j-m|=b$. This would correspond, as well known, to a linear resistive element. Otherwise, each row from $k_0$ to $2b$ contains only two nonzero elements at a distance $|j-m|\ne b$. In the example represented by eqn. \eqref{eq:algebra_2} we have assumed only two rows of this kind. The elements $y_k$ in the rows $k = 2b-1$ and $k = 2b$ are the control coefficients of the corresponding dependent sources placed in the $b-1$ and $b$ branches, respectively. Their column position specifies the kind and branch location of the control variables. Let's assume that the nonzero element is placed in column $m$. This implies that for $m<b$ the control variable is the $m$-branch current, while if $m>b$ the control variable is the $(m-b)$-branch voltage.

Therefore $\mathcal{L}$ \emph{depends} on such coefficients and hence it will be represented as $\mathcal{L} = \mathcal{L}(y_k)$.
Let us now define the matrix $\mathcal{L}_0 = \mathcal{L}(y_k = 0)$. In other words, $\mathcal{L}_0$ is obtained by setting \emph{all} the controlling coefficients $y_k$ to zero. Moreover, assume also $\Omega_{k;m}$ a $(2b\times 2b)$ matrix with all elements\footnote{It is important to stress that $k\ge k_0$ and $k_0>b+1$} $\omega_{i j} = 0, \,\, \forall i\ne k, j\ne m \,\,\, \text{and}\,\, \omega_{km} = 1$. As specified above, the $m$-index specifies the kind and position in the network of the control variable. These preliminary assumptions allow to transform eqn.~\eqref{eq:algebra_1} into the following:
\begin{eqnarray}\label{eq:algebra_3}
\left[ \mathcal{L}_0 - \sum\limits_{k = k_0}^{2b}y_k\Omega_{k;m}\right] x = U,\\
\label{eq:algebra_3bis}
\mathcal{L}_0 x = U + \sum\limits_{k = k_0}^{2b}y_k\Omega_{k;m} x.
\end{eqnarray}

Assume now that :

\begin{eqnarray}
x^{(0)} &=& \mathcal{L}_0^{-1} U, \label{eq:algebra_4}\\
x^{(k-k_0+1)} &=& y_k \mathcal{L}_0^{-1} \Omega_{k;m} x,\,\,\, k = k_0,...,2b  \label{eq:algebra_4_2}
\end{eqnarray}
and sum up, side by side, the equations:
\begin{eqnarray}\label{eq:algebra_5}
x^{(0)} + \sum\limits_{k = k_0}^{2b} x^{(k-k_0+1)} = \mathcal{L}_0^{-1} \left[ U + \sum\limits_{k = k_0}^{2b} y_k \Omega_{k;m} x \right], \\
\label{eq:algebra_6}
 \mathcal{L}_0 \left(x^{(0)} + \sum\limits_{k = k_0}^{2b} x^{(k-k_0+1)} \right) = U + \sum\limits_{k = k_0}^{2b} y_k \Omega_{k;m} x .
\end{eqnarray}
Comparing  now eqn.\eqref{eq:algebra_6} with eqn.\eqref{eq:algebra_3bis} the following condition holds:
\begin{equation}\label{eq:algebra_7}
x = x^{(0)} + \sum\limits_{k = k_0}^{2b} x^{(k-k_0+1)}.
\end{equation}
Eqn. \eqref{eq:algebra_4} represents the contribution due to the independent sources \emph{inside} the column vector $U$. Such contribution could be, in turn, split into more contributions, for each independent source in the circuit, according to the usual application of the Superposition Principle. The corresponding sub-circuit is shown in Fig.~\ref{fig:general_circuit_2}. Conversely, eqn. \eqref{eq:algebra_4_2} can re-arranged as:
\begin{equation}\label{eq:algebra_4bis}
\mathcal{L}_0 x^{(k-k_0+1)} = y_k \Omega_{k;m} x,
\end{equation}
that is, the contribution due to the controlled sources can be interpreted as the solution of the sub-circuits each fed by the voltage or current source $y_k\,x_m$. The coefficient $y_k$ is a constant representing the controlling parameter of the dependent source, while $x_m$ is the control variable that, in this frame, should be treated as an \emph{imposed parameter}, being the unknown $x^{(k-k_0+1)}$. This implies that such term could be interpreted in the corresponding circuit as an independent source on the $k-b$ branch supplying the (voltage or current) value $y_k x_m$. Fig.~\ref{fig:general_circuit_3} shows the corresponding sub-circuit. In summary:
\\\\
\begin{small}
\textsl{Provided a circuit with $M$ \emph{dependent sources}, the complete solution can be represented by adding to the contribution related to the \emph{independent sources}, $M$ further terms due to the \emph{dependent sources} as follows: the $(k-k_0+1)-$th contribution to the solution is due to the $k-$th controlled source in the $(k-b)-$th branch, treated as an independent source supplying the (current or voltage) $y_{k}\,x_m$.
}\end{small}
%
\begin{figure}[h]
\centering
\includegraphics[width=6.5cm]{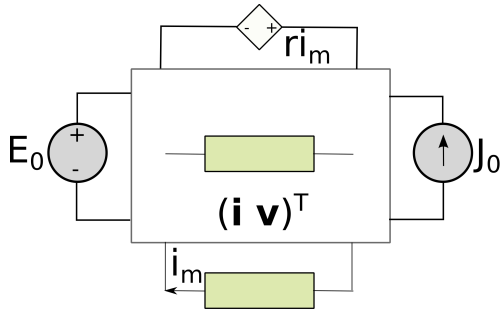}
\caption{Circuit corresponding to the algebraic system in eqn. \eqref{eq:Tableau_CS}.}
\label{fig:CS1_ex}
\end{figure}

In order to clarify the discussion, let us consider the linear resistive circuit shown in Fig. \ref{fig:CS1_ex} where, for the sake of easiness, one independent voltage source $E_0$, one independent current source  $J_0$ and one current controlled voltage source (CCVS) have been considered. In this specific case, the circuit equations in eqn.~\eqref{eq:algebra_1} take the form:

\begin{equation}\label{eq:Tableau_CS}
\begin{bmatrix}
 &   &  & & & & \\
 & \textbf{\large A} & & {\color{red}\vdots} &  & \textbf{\large 0} &  \\
 & & & {\color{red} \vdots}  & & \\
 & \textbf{\large 0} & &{\color{red}\vdots} & & \textbf{\large B} & \\
 & & &{\color{red}\vdots} & & &  \\
\hdotsfor{7} \\
 0 & \dots  0& -R \hdots & {\color{red}\vdots} & 0 & \hdots & 1 \hdots 0 \\
& \hdotsfor{6} &\\
0 & 0 &\hdots & {\color{red}\vdots} & 0 & \dots 1 & 0 \hdots\\
0 & \hdots  & 1  \hdots & {\color{red}\vdots} & 0 & 0 \hdots & 0 \hdots \\
0 & -r & \dots 0 &{\color{red}\vdots} & 0 & 0 \hdots  & 1 \hdots\\
 \end{bmatrix}
\mathbf{x} = 
\begin{bmatrix}
 0  \\
 0 \\
 \vdots \\
 \vdots \\
0 \\
E_0 \\
J_0 \\
0  \\
 \end{bmatrix}
\end{equation}
\\

Here $k_0 = 2b$, meaning that only one controlled source is present in the circuit, and further $y_{2b} = r$.
%
\begin{figure}[h]
\centering
\includegraphics[width=5.5cm]{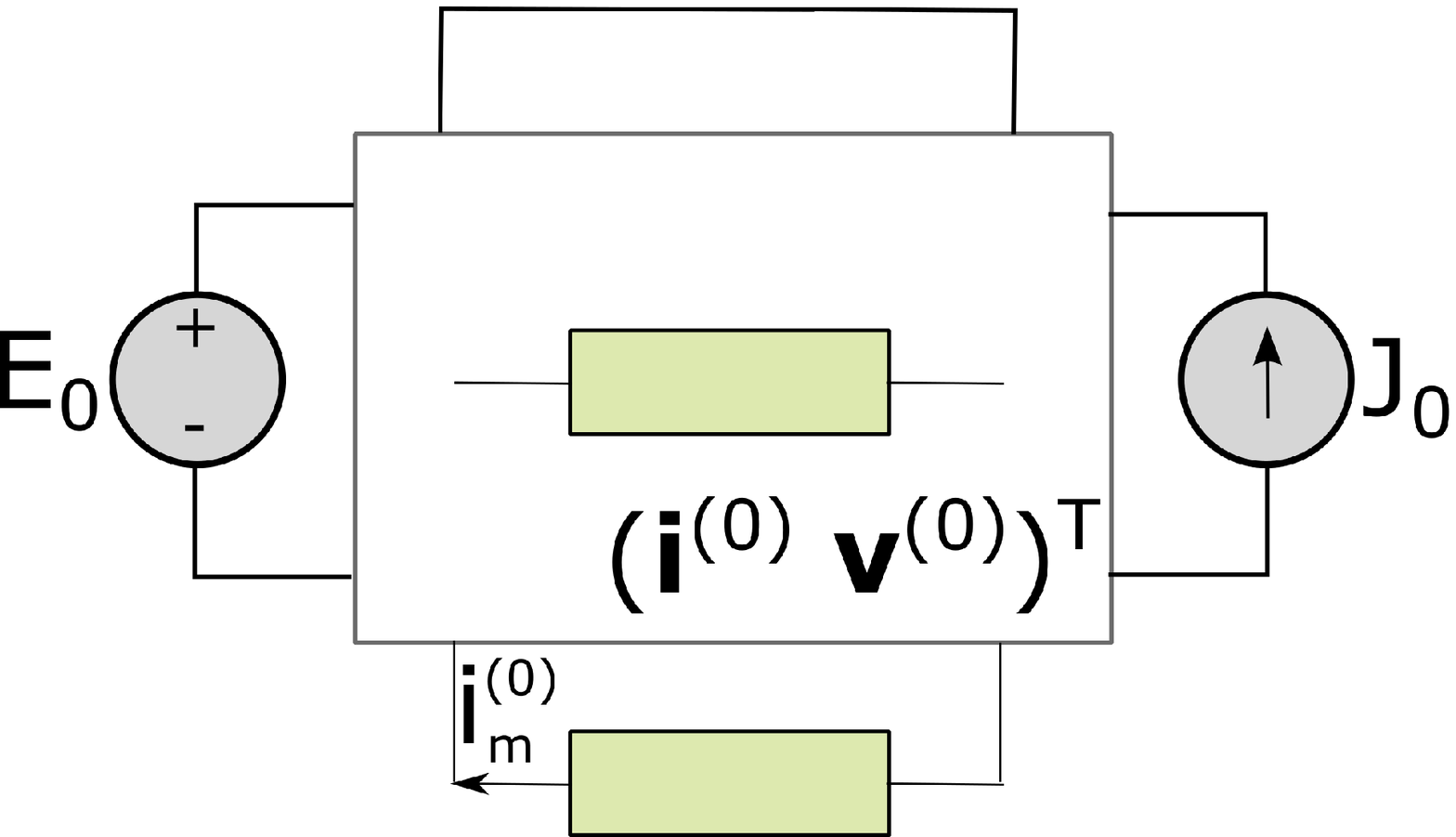}
\includegraphics[width=4.0cm]{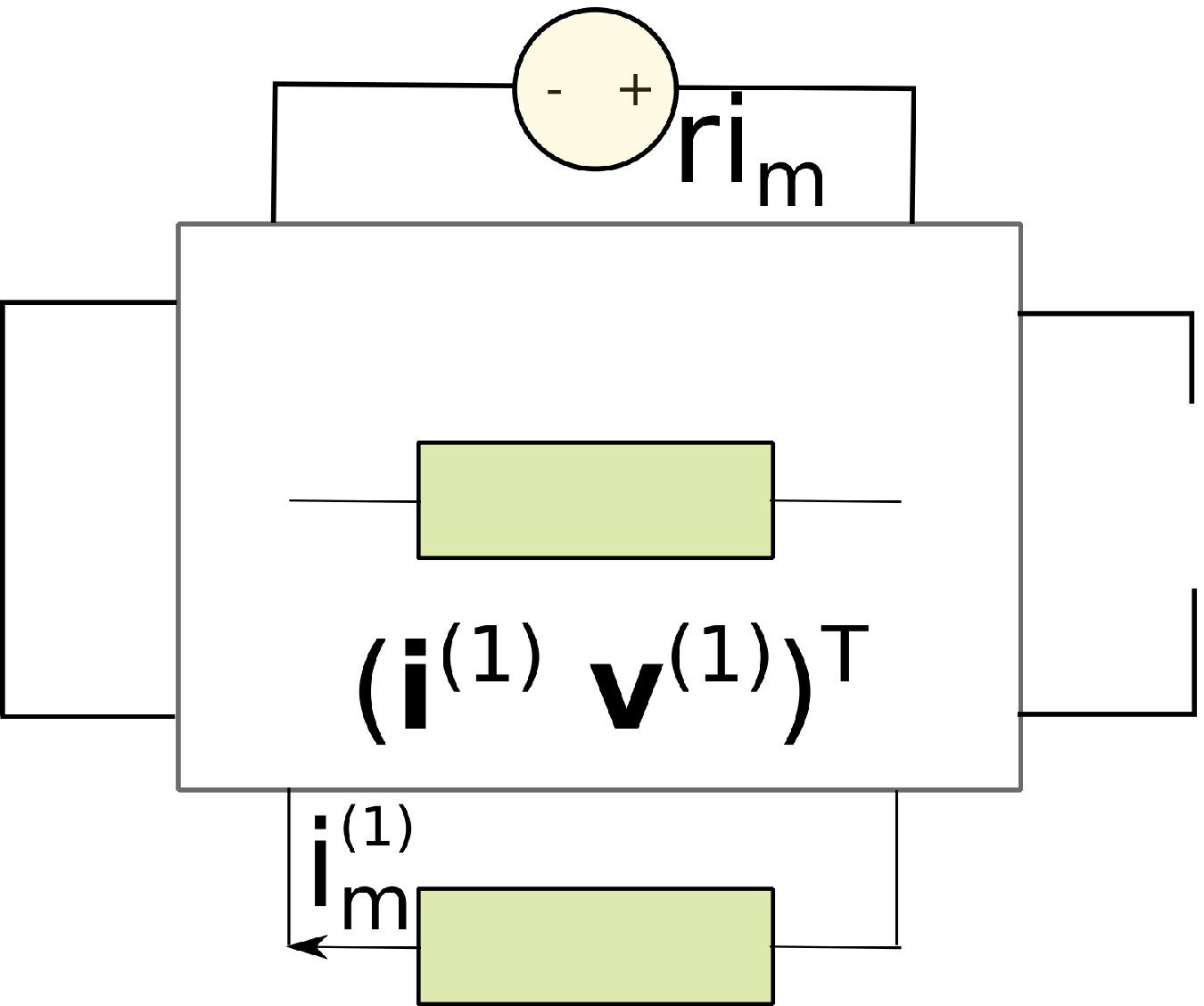}
\caption{Sub-circuits corresponding to the algebraic systems in eqns. \eqref{eq:sup_mod_3}-\eqref{eq:sup_mod_4}: a)  the dependent source is turned off; b) the independent sources are turned off.}
\label{fig:CS1_ex_2}
\end{figure}

It can be easily verified that the last row represents the CCVS, while the voltage and current sources have been shown in the two rows just above. Let us recall that for the resistor the column distance of the nonzero elements equals the number of branches, $b$. Conversely, for the controlled source, a similar situation can be observed, but the column distance of the \emph{nonzero elements} may have $|k-j| \ne b$. 
Finally, the independent sources are represented by just one nonzero element. The known terms $E_0$ or $J_0$ are in the known column vector at the rows corresponding to the specified branches.\\

If now we move the term representing the control element in the dependent source at the right hand side of equation, the known column vector could be rearranged as follows:
\begin{equation}\label{eq:new_known_term}
\mathbf{U} = 
\begin{bmatrix}
 0  \\
 0 \\
 \vdots \\
 \vdots \\
0 \\
E_0 \\
J_0 \\
0  \\
 \end{bmatrix}
 +
\begin{bmatrix}
 0  \\
 \vdots \\
 ri_m\\
 \vdots \\
0 \\
0 \\
0 \\
0\\
 \end{bmatrix}
 = \mathbf{U}_0 + r \Omega_{k;m} x
\end{equation}
where $i_m$ is the controlling current flowing in the $m$-th branch, while the coefficient matrix modifies in the following, as discussed above:

\begin{equation}\label{eq:mod_L}
\mathcal{L}_0 =
\begin{bmatrix}
 &   &  & & & & \\
 & & &  &  & &  \\
\hdotsfor{7} \\
\hdotsfor{7}\\
0 & 0 & \dots {\color{red}0} &  & 0 & 1 \hdots  & 0 \hdots \\
 \end{bmatrix} .
\end{equation}


The system:

\begin{equation}\label{eq:sup_mod_1}
\mathcal{L}_0\mathbf{x} = \mathbf{U}
\end{equation}

is linear and the solution can be written as $\mathbf{x} = \mathbf{x}^{(0)} + \mathbf{x}^{(1)}$, where the vectors

\begin{equation}\label{eq:sup_mod_2}
\mathbf{x}^{(0)} = 
\begin{bmatrix}
 i^{(0)}_1  \\
 i^{(0)}_2 \\
 \vdots \\
 i^{(0)}_b \\
v^{(0)}_1 \\
v^{(0)}_2 \\
\vdots \\
v^{(0)}_b \\
 \end{bmatrix},
 \,\,\,\,\,\,\,\,
 \mathbf{x}^{(1)} = 
 \begin{bmatrix}
 i^{(1)}_1  \\
 i^{(1)}_2 \\
 \vdots \\
 i^{(1)}_b \\
v^{(1)}_1 \\
v^{(1)}_2 \\
\vdots \\
v^{(1)}_b \\
 \end{bmatrix},
\end{equation}

satisfies the linear systems:

\begin{eqnarray}\label{eq:sup_mod_3}
\mathcal{L}\mathbf{x^{(0)}} &=& \mathbf{U^{(0)}};\\
\mathcal{L }\mathbf{x}^{(1)} &=& \mathbf{U}^{(1)}\equiv r \Omega_{k;m} x. \label{eq:sup_mod_4} 
\end{eqnarray}

It should be noticed that the column vector $x$ at the right hand side of equation contains the term $i_m$ \emph{and not} $i^{(1)}_m$ which is one of the system unknowns. In this respect, $\mathbf{U}^{(1)}$ could be fully considered as a known term.


It should appear quite evident that this line of reasoning can be extended to an arbitrary number of independent and controlled sources, provided the ordinary assumptions guaranteeing existence and uniqueness of the solution are satisfied.

It is now clear that the system in eqn. \eqref{eq:sup_mod_3} describes a network supplied by only the \emph{independent} sources. There, the controlled source has been switched off, as sketched in Fig.~ \ref{fig:CS1_ex_2}-a). The one in eqn. \eqref{eq:sup_mod_4} models the circuit in Fig. \ref{fig:CS1_ex_2}-b), where the term $ri_m$ \emph{is not} one of the circuit variables rather represented as $i^{(1)}_m$ or $v^{(1)}_m$. Hence it can be \emph{formally} considered as a known term or, in other words, an independent voltage source. In this respect the solution appears as the superposition of the response corresponding to the sub-circuit with controlled sources switched off and the response of the sub circuit with all independent sources switched off.

The above discussion shows that the \emph{apparently meaningless} handling of the system \eqref{eq:algebra_1}-\eqref{eq:algebra_2}, yields to an alternative superposition rule, as in eqns. \eqref{eq:algebra_7} and specified by eqns. \eqref{eq:algebra_4}-\eqref{eq:algebra_4_2} and \eqref{eq:sup_mod_3}-\eqref{eq:sup_mod_4}.  The latter equations correspond to the sub-circuits, shown in Fig.~\ref{fig:CS1_ex_2}. This would  show that the approach proposed so far mathematically justifies the `formal' technique, that extends the classical superposition principle when linear controlled sources are concerned.

\section{Examples}
A lot of meaningful examples have been provided in the original paper by Leach, \cite{M-L-94} and in \cite{Damper_11,Rathore_12} . So in this section we just give some few further example in order to complete the discussion. Among these we would also show how the application of superposition allows to simplify the computational effort when the Thevenin equivalent circuit or any general topological reduction is applied. \\
\subsubsection*{Example 1: VCVS controlled by an independent voltage source}
\begin{figure}[t]
\centering
\includegraphics[width=7cm]{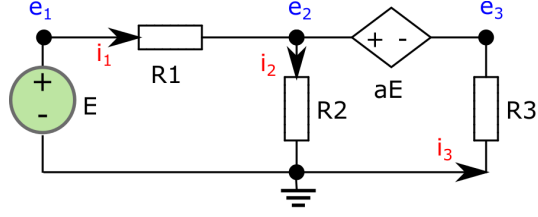}
\caption{Example 1: Circuit with one voltage controlled voltage source (VCVS) and one independent voltage source.}
\label{fig:VCVS}
\end{figure}

Let us start referring to the circuit in Fig.~\ref{fig:VCVS} of which the computation of $i_3$ is required. By adopting the node analysis, the following equations are derived:
\begin{eqnarray}
G_1 e_1 - (G_1+G_2)e_2 - G_3e_3 = 0, \\
e_1 = E,\\
e_2 - e_3 = a E,
\end{eqnarray}
that, after some algebra yield to:
\begin{eqnarray}
e_3 = \frac{G_1-a(G_1+G_2)}{G_1+G_2+G_3} E,
\end{eqnarray}

and, finally,
\begin{eqnarray}\label{eq:sol_ex_1}
i_3 = -\frac{G_1-a(G_1+G_2)}{G_1+G_2+G_3} G_3E.
\end{eqnarray}
%
%
\begin{figure}[t]
\centering
\includegraphics[width=7cm]{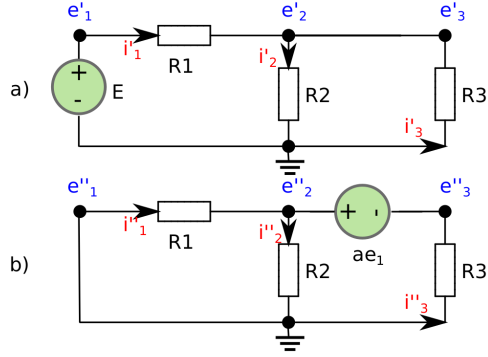}
\caption{Example 1: The sub-circuit deriving from the one in Fig.\ref{fig:VCVS} after the application of the superposition. In a) we observe the circuit with the VCVS switched off. In b) the VCVS has been replaced, according to what exposed in the preceding sections, by an independent voltage source $ae_1 = aE$.}
\label{fig:VCVS-sub}
\end{figure}
The same result can be reached by applying the superposition, as described in the previous section. To this end, let us refer to the sub-circuits sketched in Fig.~\ref{fig:VCVS-sub}. In the first circuit, the controlled source has been turned off and only the independent voltage source is there. Such a circuit allows to easily derive the contribution to the current $i_3$:
\begin{equation}
i'_3 = \frac{\frac{1}{G_2+G_3}}{\frac{1}{G_1}+\frac{1}{G_2+G_3}}G_3E,
\end{equation}
which can easily be re-arranged as:
\begin{equation}
i'_3 = - \frac{G_1}{{G_1}+{G_2+G_3}}G_3E.
\end{equation}
The second circuit is obtained by switching off the independent voltage source and observing that $e_1$ is not one of the unknown variables of the circuit and therefore can be formally considered as an independent source with value $ae_1$. At this stage therefore, the analysis is almost trivial and allows to derive the second contribution to $i_3$:
\begin{equation}
i''_3 = \frac{a e_1}{\frac{1}{G_1+G_2}+\frac{1}{G_3}},
\end{equation}
or
\begin{equation}
i''_3 = \frac{a(G_1+G_2)}{G_1+G_2+ G_3}G_3E,
\end{equation}

since $e_1 = E$. By summing up the contributions $i'_3$ and $i''_3$ the solution derived by means of the canonical procedure, as in eqn. \eqref{eq:sol_ex_1} is obtained.

\begin{figure}[h]
\centering
\includegraphics[width=6cm]{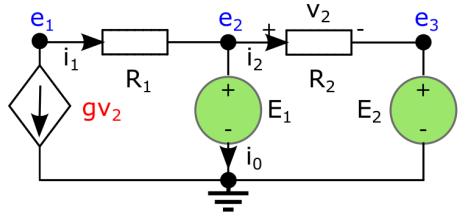}
\caption{Example 2: Circuit with one voltage controlled current source (VCCS) and two independent voltage sources.}
\label{fig:VCCS}
\end{figure}
%
\begin{figure}[h]
\centering
\includegraphics[width=7.6cm]{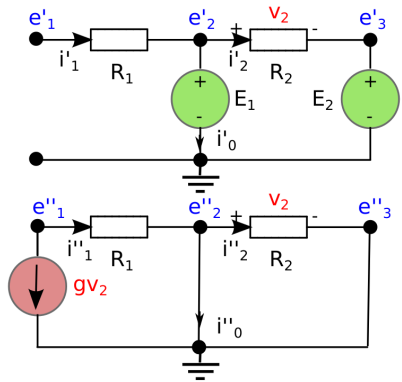}
\caption{Example 2: Sub-circuits deriving from the application of superposition to the circuit in Fig. \ref{fig:VCCS} .}
\label{fig:VCCS-sub}
\end{figure}
\subsubsection*{Example 2: VCCS and two independent sources.}
This example concerns with the circuit sketched in Fig. \ref{fig:VCCS}, where the computation of the current $i_0$ is requested. By exploiting again the standard node analysis, we can write:
\begin{eqnarray}
(e_1-e_2)G_1 &=& i_0 + (e_2-e_3)G_2; \\
e_2 &=& E_1;\\
e_3 &=& E_2;\\
G_1(e_1-e_2) &=& -g(e_2-e_3).
\end{eqnarray}
After a few algebra, the solution is drawn:
\begin{eqnarray}
i_0 = (g+G_2) (E_2-E_1).
\end{eqnarray}
Now we apply the superposition by solving the two sub-circuits shown in Fig. \ref{fig:VCCS-sub}. The contributions of the circuits to $i_0$ can be written as:
\begin{eqnarray}
i'_0 = (E_2-E_1)G_2;\\
i''_0 = -gv_2,
\end{eqnarray}
and
\begin{eqnarray}
v'_2 = (E_1-E_2);\\
v''_2 = 0.
\end{eqnarray}
By summing up we obtain:
\begin{eqnarray}
v_2 = v'_2+v''_2 &=& (E_1-E_2);\\
i_0 = i'_0+i''_0 &=& (E_2-E_1)G_2-gv_2\\
&=& (g+G_2)(E_1 - E_2),
\end{eqnarray}
which is the solution already obtained by standard methods.\\
\subsubsection*{Example 3: Thevenin equivalent and two dependent sources.}
%
%
\begin{figure}[h]
\centering
\includegraphics[width=7cm]{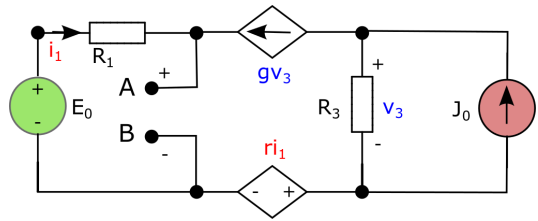}
\caption{Example 3: Circuit with one VCCS, one CCVS and two independent sources.}
\label{fig:VCCS-CCVS}
\end{figure}
%
\begin{figure}[h]
\centering
\includegraphics[width=7cm]{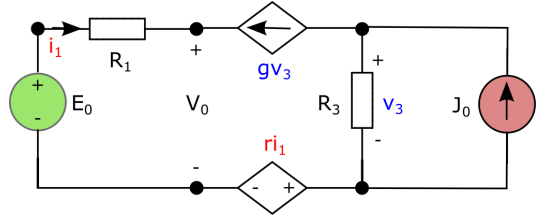}
\caption{Example 3: derivation of the $V_0$ open circuit voltage at the terminals.}
\label{fig:VCCS-CCVS-bis}
\end{figure}
%
\begin{figure}[h]
\centering
\includegraphics[width=7cm]{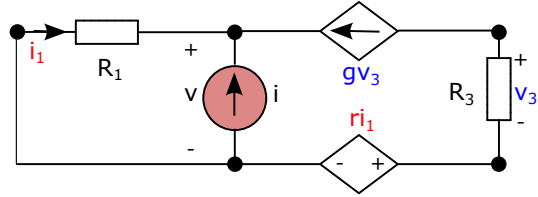}
\caption{Example 3: derivation of the $R_{eq}$ at the terminals.}
\label{fig:VCCS-CCVS-ter}
\end{figure}
As a final example, let us consider the element with two terminals shown in Fig. \ref{fig:VCCS-CCVS}, including also two controlled sources, that we aim to characterize by the aid of Thevenin Equivalent. It is trivial to realize that such circuit falls within the limits of the discussions provided above that justifies the applicability of the superposition also in this case. The circuit in Fig.~\ref{fig:VCCS-CCVS} represents a specific example that we analyze in detail by the aid of the superposition tool, leaving to the reader the checking by the standard node analysis. Let us assume that the Th\'evenin equivalent at the terminals $AB$ is required. To this aim, the open circuit voltage and the equivalent resistance at the terminals are needed. The circuit shown in Fig.~\ref{fig:VCCS-CCVS-bis} shows the $V_0$ open circuit voltage to be calculated. To this aim four sub-circuits can be considered (not shown for shortness), each with only a source switched on. It is quite trivial to compute the contributions to $v_3$ and $V_0$ voltages, as follows:
\begin{eqnarray}
E_0 \,\, \text{`on':}\,\,\, \,\,\,V^{(1)}_0 = E_0;\\
v^{(1)}_3 = 0;
\end{eqnarray}
\begin{eqnarray}
gv_3 \,\, \text{`on':}\,\,\,\,\,\,V^{(2)}_0 = \frac{g}{G_1}v_3;\\
v^{(2)}_3 = -\frac{g}{G_3}v_3;
\end{eqnarray}
\begin{eqnarray}
ri_1 \,\, \text{`on':}\,\,\,\,\,\,V^{(3)}_0 = 0;\\
v^{(3)}_3 = 0;
\end{eqnarray}
\begin{eqnarray}
J_0 \,\, \text{`on'}\,\,\,\,\,\,V^{(4)}_0 = 0;\\
v^{(4)}_3 = \frac{J_0}{G_3}.
\end{eqnarray}
By summing up, the following result is drawn:
\begin{eqnarray}
v_3 &=&v^{(1)}_3+v^{(2)}_3+v^{(3)}_3 + v^{(4)}_3 = \frac{J_0-gv_3}{G_3},\\
V_0 &=& V^{(1)}_0 + V^{(2)}_0 + V^{(3)}_0 + V^{(4)}_0 = E_0 + \frac{gv_3}{G_1},\\
\end{eqnarray}
or, equivalently: \\
\begin{eqnarray}
v_3 &=& \frac{J_0}{\left(1 + \frac{g}{G_3}\right)};\\
V_0 &=& E_0 + \frac{g/G_1}{\left(1 + \frac{g}{G_3}\right)}J_0.
\end{eqnarray}
Referring now to the circuit in Fig.~\ref{fig:VCCS-CCVS-ter}, the equivalent resistance is given by $R_{eq} = \frac{v}{i}$. As before, by solving the three sub-circuits each with only one source `on' at the time, the following equations are derived:
\begin{eqnarray}
i \,\, \text{`on':}\,\,\, \,\,\,v^{(1)} = \frac{i}{G_1};\\
v^{(1)}_3 = 0;
\end{eqnarray}
\begin{eqnarray}
gv_3 \,\, \text{`on':}\,\,\,\,\,\,v^{(2)} = \frac{g}{G_1}v_3;\\
v^{(2)}_3 = -\frac{g}{G_3}v_3;
\end{eqnarray}
\begin{eqnarray}
ri_1 \,\, \text{`on':} \,\,\,\,\,\,V^{(3)}_0 = 0;\\
v^{(3)}_3 = 0;
\end{eqnarray}
By summing up
\begin{eqnarray}
v_3 &=& - \frac{g}{G_3}v_3 \,\, \Rightarrow v_3 = 0;\\
V &=& \frac{i}{G_1} + \frac{g}{G_1}v_3 \,\, \Rightarrow R_{eq} = 1/G_1 = R_1.
\end{eqnarray}
This should clearly show the applicability of the Superposition Principle when controlled sources are of concern.

\section{Conclusion}
The discussion carried out in the paper answers a question arisen in my teaching activity. It followed the ideas by W.~M. Leach, \cite{M-L-94} and confirmed by R.~I.Damper, \cite{Damper_11} and  T.~S. Rathore, \cite{Rathore_12} in later contributions. In the manuscript it is demonstrated, through a suitable interpretation of the turning on/off notion of a controlled source, that all the tools usually adopted in the network analysis by the Superposition Principle still works when linear controlled sources are concerned. The manuscript deviates from the original derivation by proposing a more formal and general approach that still holds the advantage to be straightforward and suitable for the needs of an undergraduate class in Electrical Engineering.

\end{document}